\documentclass{epsconf}
\usepackage{graphicx}
\usepackage{wrapfig}
\usepackage{mathrsfs}
\usepackage{amsmath}

\title{Core localized alpha-channeling via low frequency Alfv\'en mode generation in reversed shear scenarios}
\author{Z. Qiu$^{1,2}$, S. Wei$^1$, T. Wang$^{1,2}$, L. Chen$^{1, 3}$ and F. Zonca$^{2,1}$}
\institute{$^1$Inst.    Fusion Theory \& Simulation, School of Physics, Zhejiang Univ., Hangzhou, P.R.C.\\
$^2$Center for Nonlinear Plasma Science and C.R. ENEA Frascati, C.P. 65, 00044 Frascati, Italy\\
$^3$Dept.   Physics \& Astronomy,  University of California, Irvine CA 92697-4575, U.S.A.}
\begin{document}
\maketitle

\begin{abstract}
A novel channel for fuel ions heating in tokamak core plasma is proposed and analyzed using nonlinear gyrokinetic theory. The channel is achieved via spontaneous decay of reversed shear Alfv\'en eigenmode (RSAE) into low frequency  Alfv\'en  modes (LFAM), which then heat  fuel ions via collisionless ion Landau damping.  The conditions for RSAE spontaneous decay are investigated, and the saturation level and the consequent fuel ion heating rate are also derived.   The channel is expected to be crucial for future reactors operating under reversed shear configurations, where fusion alpha particles are generated in the tokamak core where the magnetic shear is typically reversed, and there is a dense RSAE spectrum due to the small alpha particle characteristic dimensionless orbits.
\end{abstract}

Energetic particles (EPs) as well as   fusion alpha particles related physics \cite{AFasoliNF2007,LChenRMP2016} are key elements  towards understanding  the performance of future fusion reactors, among which two crucial topics are    EPs   transport loss by self-generated   collective oscillations  such as shear Alfv\'en wave (SAW)  eigenmodes  \cite{AFasoliNF2007,LChenRMP2016} and searching for alternative/complementary routes to transfer EP power to fuel ions,  i.e.,  alpha-channeling \cite{NFischPRL1992}. Both processes are  influenced  by the saturation level and spectrum of SAWs.   In this contribution, a  channel  for reversed shear Alfv\'en eigenmode (RSAE) \cite{HBerkPRL2001} nonlinear saturation is proposed and analysed, which is expected  to play significant roles in future reactor-scale tokamaks with  rich spectrum of core-localized RSAEs \cite{TWangPoP2018} due to the reversed shear magnetic configuration   and small dimensionless  EP   orbit size.  
In this proposed process, a RSAE spontaneously decays into another RSAE and a low frequency Alfv\'en  mode (LFAM), which can be ion  Landau damped, leading to effective heating of thermal ions in the reversed shear region, and consequently, enhanced fusion performance.

We consider for simplicity  low-$\beta_i$ plasma such that the frequency separation between RSAE and LFAM required for resonant mode coupling can be well satisfied. The nonlinear coupling is dominated by thermal plasma contribution, while the RSAEs are excited by EPs, so the thermal plasma nonuniformity can be neglected, which is also consistent with the advanced scenario of reversed shear configuration. 
The governing equations describing nonlinear interactions among RSAEs and LFAM with all predominantly SAW polarization  can be derived from nonlinear gyrokinetic vorticity equation \cite{LChenJGR1991} 
 and   quasi-neutrality condition, 
with the particle response  derived from nonlinear gyrokinetic equation \cite{EFriemanPoF1982}.

The general equation for three SAWs nonlinear interaction, with the matching condition being $\Omega_3(\omega_3,\mathbf{k}_3)=\Omega_1(\omega_1,\mathbf{k}_1)+\Omega_2(\omega_2,\mathbf{k}_2)$, can be derived as\small
\begin{eqnarray}
b_{k_3}\mathcal{E}_{k_3}\delta\phi_{k_3} &=&- \frac{i}{\omega_3} \Lambda^{k_3}_{k_2,k_1} \left[  (b_{k_2}-b_{k_1})\left(1-\frac{k_{\parallel1}k_{\parallel2}V^2_A}{\omega_{1}\omega_{2}}\right)  +b_{k_3}V^2_A\frac{k_{\parallel 3}}{\omega_3}\left(\frac{k_{\parallel 1}}{\omega_{1}} - \frac{k_{\parallel 2}}{\omega_{2}}\right)  \right]\delta\phi_{k_1}\delta\phi_{k_2},\label{eq:NL_SAW}
\end{eqnarray}\normalsize
with   $\mathcal{E}_k\equiv   -k^2_{\parallel} V^2_A/\omega^2_k  + 1 - \omega^2_G/\omega^2_k $ being the SAW dielectric function in the WKB limit,  $\omega_G\equiv \sqrt{7/4+T_e/T_i} v_i/R_0$ being the leading order geodesic acoustic mode frequency \cite{NWinsorPoF1968}, accounting for SAW continuum upshift and creation of beta-induced continuum gap, and $\Lambda^k_{k'',k'}\equiv (c/B_0)\hat{\mathbf{b}}\cdot\mathbf{k''}\times\mathbf{k'}$ with $\hat{\mathbf{b}}$ being the unit vector along the equilibrium magnetic field $\mathbf{B}_0$.  

Equation (\ref{eq:NL_SAW}) describes the nonlinear evolution of SAWs,  with $\Omega_3$  modified by the beating of $\Omega_1$ and $\Omega_2$,   the first term on the right hand side  due to  the competition of Reynolds and Maxwell stresses, and the second term from finite parallel electric field contribution to field line bending. Note that,  since $(\omega_1+\omega_2)\simeq (k_{\parallel 1}+k_{\parallel 2})V_A$,     $ \Omega_3$  naturally satisfies the SAW D.R.  and can be strongly excited  if it is a normal mode of the system,  leading to  significant spectral transfer  of SAW turbulence.   
We note that, in the expression of $\mathcal{E}_k$, effects of wave-particle interactions are not included,   consistent with the $k_{\parallel}v_i\ll\omega_k$ ordering for bulk non-resonant ions.  However, finite Landau damping due to resonance with ions  is crucial for alpha-channeling, and will be recovered formally in the later analysis by inclusion of the anti-Hermitian part of $\mathscr{E}_k$ \cite{FZoncaPPCF1996}.

\section{Parametric decay of RSAE}
\label{sec:NL_DR}

Equation (\ref{eq:NL_SAW}) will be applied to the nonlinear decay of a pump RSAE   $\Omega_0(\omega_0, \mathbf{k}_0)$  into a RSAE sideband  $\Omega_1(\omega_1, \mathbf{k}_1)$ and a LFAM  $\Omega_B(\omega_B, \mathbf{k}_B)$, with the frequency/wavenumber matching condition $\Omega_0=\Omega_1+\Omega_B$ assumed without loss of generality.
For RSAE and LFAM being dominated by single-$n$ and single-$m$ mode structures, we take
 $\delta\phi_k=A_k(t)\Phi_k(x) \exp{\left(-i\omega_k t+in\xi-im\theta\right)}$, with  $A_k(t)$ being the slowly varying mode amplitude, $\Phi_k(x)$ the parallel mode structure localized about $q_{min}$ with $x\equiv nq-m$, and    the normalization condition $\int |\Phi_k|^2 dx=1$ is satisfied.
 For the effective transfer of alpha particle energy to core ions, $\omega_B\leq O(v_i/(qR_0))$, and thus, $|\omega_B|\ll |\omega_0|, |\omega_1|$ and $k_{\parallel B}\simeq 0$. Thus, the $q_{min}$ surface  also corresponds to the rational surface of $\Omega_B$, i.e., $\Omega_B$ is the LFAM in the reversed shear configuration, as investigated   theoretically \cite{RMaPPCF2022}.  We then have, $\omega_0\simeq\omega_1$ and $k_{\parallel0}\simeq k_{\parallel 1}$.  Effects of small frequency mismatch on the decay process will be discussed later.

The nonlinear RSAE  sideband  and LFAM equations can be derived  from equation (\ref{eq:NL_SAW}) as
\begin{eqnarray}
\hat{b}_1\hat{\mathcal{E}}_1 A_1 &=&-\frac{i}{\omega_1} \left\langle \Lambda^{k_1}_{k_0,k_{B^*}}\alpha_1 \Phi_1\Phi_0\Phi_{B}\right\rangle_x A_0 A_{B^*}, \label{eq:RSAE_sideband_eigen}\\
\hat{b}_B\hat{\mathcal{E}}_B A_B&=&-\frac{i}{\omega_B}  \left\langle \Lambda^{k_B}_{k_0,k_{1^*}}\alpha_B \Phi_B\Phi_0\Phi_{1}\right\rangle_x A_0 A_{1^*}, \label{eq:LFAM_eigen}
\end{eqnarray}
with  $\alpha_1\equiv (b_0-b_B)(1- k_{\parallel B}k_{\parallel0}V^2_A/(\omega_0\omega_B))  +  b_1 V^2_A  (k_{\parallel 1}/\omega_{1} ) (k_{\parallel B}/\omega_{B} - k_{\parallel0}/\omega_{0})$, $\alpha_B \equiv (b_0-b_1)(1- k_{\parallel 1}k_{\parallel0}V^2_A/(\omega_0\omega_1))  +  b_B V^2_A  (k_{\parallel B}/\omega_{B} ) (k_{\parallel 1}/\omega_{1} - k_{\parallel0}/\omega_{0})$, $\langle \cdots\rangle_x\equiv \int \cdots dx$ denoting   averaging over the fast radial scale,   $\hat{b}_1\hat{\mathcal{E}}_1\equiv \int \Phi_1 b_1 \mathcal{E}_1\Phi_1 dx$ being the $\Omega_1$ eigenmode  local dispersion function, and $\hat{b}_B\hat{\mathcal{E}}_B$ being the local  dispersion function for the  LFAM eigenmode.

The parametric decay dispersion relation for RSAE  decaying into another RSAE and  LFAM  can then be derived  by combining equations (\ref{eq:RSAE_sideband_eigen}) and (\ref{eq:LFAM_eigen})
\begin{eqnarray}
\hat{\mathcal{E}}_1\hat{\mathcal{E}}_{B^*} \simeq  \left(\hat{\Lambda}^{k_1}_{k_0,k_{B^*}}\right)^2\frac{\hat{\alpha}_N}{\hat{b}_B\hat{b}_1 \omega_{B}\omega_{1}} \hat{C}^2 |A_0|^2, \label{eq:parametric_disp}
\end{eqnarray}
with  $\hat{C}\equiv \langle\Phi_{0}\Phi_B\Phi_1\rangle_x$,  $ \hat{\Lambda}^{k_1}_{k_0,k_{B^*}}=  \langle\Lambda^{k_1}_{k_0,k_{B^*}}\rangle_x$, $\hat{\alpha}_N\equiv \hat{\alpha}_1\hat{\alpha}_B$,   and  $\hat{C} \simeq  \sqrt{2 \Delta_B/(\sqrt{\pi}\Delta_0\Delta_1)}$, with $\Delta_0\sim\Delta_1\sim O(1)$ and $\Delta_B\sim O(\beta^{1/2})$ being the characteristic radial widths of the respective linear parallel mode structures.
Expanding $\hat{\mathcal{E}}_{1}\simeq i \partial_{\omega_1}\hat{\mathcal{E}}_{1}(\partial_t+\gamma_1)\simeq (2 i/\omega_1) (\gamma+\gamma_1)$ and $\hat{\mathcal{E}}_{B^*}\simeq (-2i/\omega_B) (\gamma+\gamma_B)$ in the local limit, with   $\gamma$ denoting the slow temporal variation of $\Omega_1$ and $\Omega_B$ due to the parametric instability, and $\gamma_1/\gamma_B$ being the linear damping  rates of RSAE/LFAM accounted for by the anti-Hermitian part of $\mathcal{E}_1/\mathcal{E}_B$,  one obtains
\begin{eqnarray}
(\gamma+\gamma_1)(\gamma+\gamma_B)=\left(\hat{\Lambda}^{k_1}_{k_0,k_{B^*}}\right)^2 \frac{\hat{\alpha}_N}{4 \hat{b}_B \hat{b}_1}  \hat{C}^2|A_0|^2. \label{eq:parametric_DR}
\end{eqnarray}
The condition for the pump RSAE spontaneous decay can thus be obtained from equation (\ref{eq:parametric_DR}) as
$\hat{\alpha}_N>0$ 
    and
$  (\hat{\Lambda}^{k_1}_{k_0,k_{B^*}})^2 \hat{\alpha}_N  \hat{C}^2|A_0|^2/(4 \hat{b}_B \hat{b}_1) >\gamma_B\gamma_1$ for the nonlinear drive overcoming the threshold due to $\Omega_1$ and $\Omega_B$ Landau damping.

The nonlinear dispersion relation is very complex, and depends on various conditions including the polarization   and mode structure  of the three modes involved.  For  further  analytical progress,    the WKB limit  and  the strong assumption of $k_{\parallel B}\rightarrow 0$ is adopted, and a  parameter regime  can be identified for the spontaneous decay process to strongly occur, which  corresponds to $k_{\perp1}\gg k_{\perp0}$, such that $(b_0-b_1)(b_0-b_B-b_1)>0$; and $\hat{\alpha}_N>0$ can be satisfied with $1-k_{\parallel0}k_{\parallel 1}V^2_A/(\omega_0\omega_1)>0$,  which generally requires  $\Omega_1$ being excited above the local SAW continuum accumulation point with $n_1q_{min}< m_1$.

The threshold condition for the RSAE spontaneous decay, for the proposed parameter region  of RSAE ``normal cascading" to $|k_{\perp1}|\gg |k_{\perp0}|$,  can be estimated as
\begin{eqnarray}
\left|\frac{\delta B_{\perp0}}{B_0}\right|^2 &>& \frac{4\gamma_1\gamma_B}{\omega_0\omega_1} \frac{k^2_{\parallel0}}{k^2_{\perp1}} \frac{1}{\hat{C}^2}  \frac{1}{1-k_{\parallel0}k_{\parallel 1}V^2_A/(\omega_0\omega_1)}               \sim  \mathcal{O}(10^{-7}),
\end{eqnarray}
and is   comparable with or slightly higher than typical threshold condition for other dominant nonlinear mode coupling processes, e.g., ZS generation. This threshold amplitude, is also  consistent with typical SAW instability intensity observed in experiments. Thus, this channel could be an important process in determining the nonlinear dynamics of RSAE.

\section{Nonlinear saturation and core-localized ion heating}
\label{sec:saturation}

The RSAE saturation level   can be estimated by considering the feedback of the two sidebands to the pump RSAE, which can be derived from equation (\ref{eq:NL_SAW}) as
\begin{eqnarray}
\hat{b}_0\hat{\mathcal{E}}_0 A_0\simeq -\frac{i}{\omega_{0}}\hat{\Lambda}^{k_0}_{k_1,k_{B}} \hat{\alpha}_0 \hat{C} A_1 A_B,\label{eq:RSAE_pump}
\end{eqnarray}
with $\alpha_0= (b_1-b_B)  (1- k_{\parallel B}k_{\parallel 1}V^2_A/(\omega_1\omega_B))  +  b_0 V^2_A(k_{\parallel0}/\omega_{0})  (k_{\parallel B}/\omega_{B}- k_{\parallel 1}/\omega_{1}) $. The saturation level of LFAM, can be estimated from the fixed point solution of   equations (\ref{eq:RSAE_sideband_eigen}), (\ref{eq:LFAM_eigen}) and (\ref{eq:RSAE_pump}), and one obtains, 
$|A_B|^2=  \gamma_0\gamma_1 \hat{b}_0\hat{b}_1\omega_0\omega_1\partial_{\omega_1}\mathcal{E}_{1,\mathcal{R}}\partial_{\omega_0}\mathcal{E}_{0,\mathcal{R}}/(\hat{\alpha}_0 \hat{\alpha}_1 |\hat{C}|^2 (\hat{\Lambda}^{k_0}_{k_1,k_B})^2)$, and the ion heating rate due to LFAM Landau damping, can be estimated as
\begin{eqnarray}
P_i=2\gamma_B \omega_B\frac{\partial \mathscr{E}_{B,\mathcal{R}}}{\partial\omega_B}\frac{n_0e^2}{T_i}\hat{b}_B |A_B|^2 \sim 10^{-3}\gamma_0 n T.
\end{eqnarray}
The obtained core ion heating due to LFAM conllisionless damping, can be comparable  to Coulomb collisional heating estimated by $n T/\tau_E$,  with $\tau_E$ being the energy confinement time. 

This channel, achieved via the Landau damping of secondary LFAM, noting that $k_{\parallel B} \ll1$, is highly localized around the $q_{min}$ surface (this conclusion can also be obtained, noting  as the ``secondary" LFAM structure will be determined by the primary RSAE, with a narrower extent than the primary RSAEs), will deposit fusion alpha particle power locally and heating core ions, leading to direct improvement of fusion performance in the tokamak center. The nonlinear dynamics of RSAE with multiple channels accounted for simultaneously \cite{TWangPoP2018,SWeiJPP2021,SWeiNF2022}  is crucial for the understanding of core plasma behaviour and fusion performance of future reactors.

\end{document}